\documentclass[
 reprint,
superscriptaddress,
 amsmath,amssymb,
 aps,
prb,
floatfix,
]{revtex4-2}

\usepackage{graphicx}% Include figure files
\usepackage{dcolumn}% Align table columns on decimal point
\usepackage{bm}% bold math
\usepackage{siunitx}

\newcommand{\dfield}{D}
\newcommand{\deps}{D/\varepsilon_0}
\newcommand{\figtext}{Fig.\@}
\newcommand{\figstext}{Figs.\@}

\renewcommand{\selectlanguage}[1]{}
\begin{document}

% SHA added to make citations nicer
\setcitestyle{super}

% SHA added to make figure titles nicer
\renewcommand\figurename{Fig.}

\title{Displacement field-controlled fractional Chern insulators and charge density waves in a graphene/hBN moir\'{e} superlattice}

\author{Samuel H. Aronson}
\altaffiliation{These authors contributed equally.}
\affiliation{Department of Physics, Massachusetts Institute of Technology, Cambridge, MA, USA}
\author{Tonghang Han}
\altaffiliation{These authors contributed equally.}
\affiliation{Department of Physics, Massachusetts Institute of Technology, Cambridge, MA, USA}
\author{Zhengguang Lu}
\affiliation{Department of Physics, Massachusetts Institute of Technology, Cambridge, MA, USA}
\author{Yuxuan Yao}
\affiliation{Department of Physics, Massachusetts Institute of Technology, Cambridge, MA, USA}
\author{Kenji Watanabe}
\affiliation{Research Center for Electronic and Optical Materials, National Institute for Materials Science, Tsukuba,
Japan}
\author{Takashi Taniguchi}
\affiliation{Research Center for Materials Nanoarchitectonics, National Institute for Materials Science, Tsukuba, Japan}
\author{Long Ju}
\altaffiliation{longju@mit.edu}
\affiliation{Department of Physics, Massachusetts Institute of Technology, Cambridge, MA, USA}
\author{Raymond C. Ashoori}
\altaffiliation{ashoori@mit.edu}
\affiliation{Department of Physics, Massachusetts Institute of Technology, Cambridge, MA, USA}

\preprint{APS/123-QED}

\begin{abstract}
Rhombohedral multilayer graphene, with its flat electronic bands and concentrated Berry curvature\cite{xiao_valley-contrasting_2007,min_electronic_2008,koshino_trigonal_2009,han_orbital_2023}, is a promising material for the realization of correlated topological phases of matter. When aligned to an adjacent hexagonal boron nitride (hBN) layer, the graphene develops narrow minibands with non-trivial topology\cite{kumar_flat_2013,zhang_nearly_2019,chittari_gate-tunable_2019,chen_tunable_2020,park_topological_2023}. By tuning an externally-applied electric displacement field, the conduction electrons can either be pushed towards or away from the moir\'{e} superlattice. Motivated by the recent observation of the fractional quantum anomalous Hall effect (FQAHE) in the moir\'{e}-distant case\cite{lu_fractional_2024}, we study the opposite moir\'{e}-proximal case, where the superlattice potential is considerably stronger. We explore the physics within the moir\'{e} conduction bands through capacitance measurements that allow us to determine the inverse electronic compressibility and extract energy gaps of incompressible states. We observe integer and fractional Chern insulator states at superlattice filling factors $\nu=1$, 2/3, and 1/3 with St\v{r}eda slopes of -1, -2/3, and -1/3, respectively. Remarkably, the $\nu=1/3$ state persists down to a magnetic field of \qty{0.2}{\tesla}. In addition, we also observe numerous trivial and topological charge density waves. We map out a phase diagram that is highly sensitive to both displacement and magnetic fields, which tune the system between various ground states by modifying the band dispersion and the structure of the electronic wavefunctions. This work demonstrates displacement field control of topological phase transitions in the moir\'{e}-proximal limit of rhombohedral pentalayer graphene, creating a highly-tunable platform for  studying the interplay between intrinsic band topology and strong lattice effects.

% We observe a variety of competing correlated phases, including both trivial and topological charge density waves as well as a fractional Chern insulator persisting to nearly zero magnetic field.

% We observe a variety of competing correlated phases, including a fractional Chern insulator at superlattice filling factor $\nu=1/3$ persisting down to a magnetic field of \qty{0.2}{\tesla}, as well as both trivial and topological charge density waves.

\end{abstract}

\maketitle

%\tableofcontents
In the presence of an external magnetic field, the continuous energy bands of a two-dimensional electron gas (2DEG) break up into discrete, highly degenerate Landau levels. When the chemical potential sits between Landau levels, the 2DEG exhibits the integer quantum Hall effect, in which the system has a bulk gap but supports dissipationless chiral edge modes and a Hall conductance quantized to an integer multiple of $e^2/h$\cite{klitzing_new_1980}. At certain rational fillings of a Landau level, electronic correlations open up a gap in the many-body spectrum, leading to the fractional quantum Hall effect (FQHE)\cite{tsui_two-dimensional_1982}. Flat, isolated Chern minibands in moir\'{e} materials are lattice analogs of the field-induced Landau levels required for conventional quantum Hall physics. In these bands, Berry curvature plays a role similar to that of an external magnetic field, while the narrow bandwidth quenches the electrons' kinetic energy, allowing electronic correlations to dominate the low-energy physics. When a Chern band is fully filled, the result is a Chern insulator (CI), a state which, like a conventional quantum Hall insulator, hosts dissipationless edge modes and quantized Hall conductance\cite{haldane_model_1988}. In analogy to correlations in a partially-filled Landau level producing a fractional quantum Hall insulator, a partially-filled Chern band can give rise to a fractional Chern insulator (FCI)\cite{tang_high-temperature_2011,sun_nearly_2011,neupert_fractional_2011,sheng_fractional_2011,regnault_fractional_2011,parameswaran_fractional_2013,bergholtz_topological_2013}.
\begin{figure*}
\includegraphics[width=\textwidth]{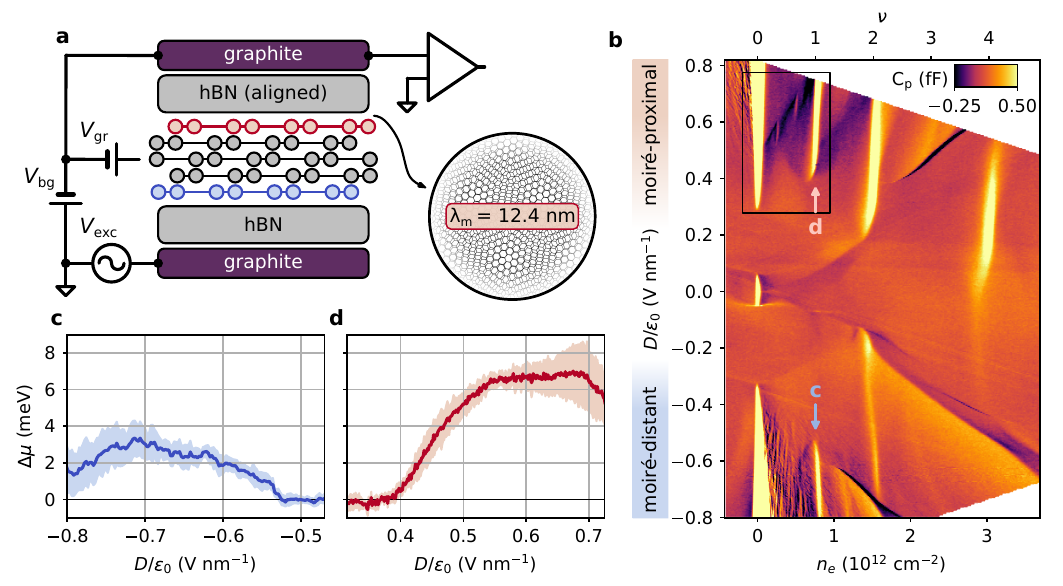}
\caption{\label{fig:fig1} \textbf{Electronic compressibility of the rhombohedral pentalayer graphene superlattice.} \textbf{a,} Penetration capacitance measurement scheme. An ac excitation $V_\text{exc}$ is applied to the bottom gate, and the penetrating signal is measured on the top gate. The effective top and bottom gate voltages are $V_\text{t}=-V_\text{gr}$ and $V_\text{b}=V_\text{bg}-V_\text{gr}$, respectively. A moir\'{e} superlattice forms at the interface between the top layer of graphene and the aligned hBN. \textbf{b,} Map of penetration capacitance measured as a function of density and displacement field at \qty{0}{\tesla}. Bright features correspond to incompressible states. Vertical streaks appearing at the valence band edge for $\dfield>0$ and conduction band edge for $\dfield<0$ result from the formation of pn junctions at the contacts, which lead to difficulties in the compressibility measurement\cite{zhou_half-_2021}. The box in the upper left corner corresponds to the area of focus in \figstext{} \ref{fig:fig2} and \ref{fig:fig3}. \textbf{c,\,d,} Widths of the moir\'{e}-distant (\textbf{c}) and moir\'{e}-proximal (\textbf{d}) $\nu=1$ gaps as a function of displacement field. A detailed analysis of the extraction of gap widths and error bands from the compressibility data is presented in the supplement.}
\end{figure*}

Even though Chern bands have intrinsic topology, an external magnetic field is often required to induce FCI ground states\cite{spanton_observation_2018,xie_fractional_2021}. However, recent experiments demonstrated the fractional quantum anomalous Hall effect (FQAHE), the zero-field counterpart to the FQHE, in two remarkably different systems. In twisted bilayer MoTe$_2$\cite{cai_signatures_2023,zeng_thermodynamic_2023,park_observation_2023}, a strong superlattice potential generates flat, isolated moir\'{e} bands which are similar to Landau levels at the single-particle level. Large spin-orbit coupling and layer pseudospin texture are essential for generating the FQAHE in this system\cite{wu_topological_2019,devakul_magic_2021}. In contrast, the FQAHE was realized without these ingredients in the rhombohedral graphene-hBN superlattice (R5G-hBN)\cite{lu_fractional_2024} when the conduction electrons were pushed \textit{away} from the moir\'{e} interface by an external electric displacement field. Under these conditions, the superlattice potential is relatively weak, and there is considerable overlap between moir\'{e} conduction bands in the single-particle picture. Electronic interactions are therefore critical for isolating the topological miniband which hosts the FQAHE, triggering debate around its underlying origin in this material\cite{dong_theory_2023,zhou_fractional_2023,dong_anomalous_2023,kwan_moire_2023,guo_theory_2023,huang_self-consistent_2024}. 

To shed light on this mystery, we study the opposite moir\'{e}-proximal case, when the displacement field pushes the conduction electrons towards the moir\'{e} interface. In this limit, the stronger superlattice potential may significantly modify the dispersion and topological structure of the minibands\cite{zhang_nearly_2019}. We address the moir\'{e}-proximal limit experimentally through capacitance measurements. While local compressibility probes, such as scanning single-electron transistors, can avoid defects and twist angle disorder, they generally do not allow for uniform and well-controlled displacement fields. Planar capacitance experiments, on the other hand, enable precise control over displacement field and are not hindered by tip gating effects. In addition, they provide quantitative measurements of gap widths and the thermodynamic density of states.

\section{Phase diagram of R5G-hBN}
We study a dual-gated device consisting of a sheet of rhombohedral pentalayer graphene encapsulated between two dielectric hBN layers. The graphene is crystallographically aligned to the hBN on the top side to produce a long-wavelength moir\'{e} superlattice potential. By tuning the effective top and bottom gate voltages $V_t$ and $V_b$, we can independently control the electron density $n_e=(c_bV_b+c_tV_t)/e$ and applied perpendicular electric displacement field $\deps=(c_bV_b-c_tV_t)/2$ where $c_t$ and $c_b$ are the geometric capacitances per unit area from the top and bottom gates to the graphene. By applying an ac voltage excitation on the bottom gate and measuring the resultant current fluctuations on the top gate (\figtext{} \ref{fig:fig1}a), we can determine the penetration capacitance $c_p$ and hence the electronic compressibility $\partial n/\partial\mu$:
\begin{equation}
    \frac{1}{c_p}=\frac{1}{c_t}+\frac{1}{c_b}+\frac{e^2}{c_tc_b}\left(\frac{\partial n}{\partial\mu}\right)
\end{equation}

\figtext{} \ref{fig:fig1}b shows the measured capacitance of an R5G-hBN device as a function of electron density and displacement field. The device has a moir\'{e} wavelength of \qty{12.4}{\nano\meter}, corresponding to a twist angle of approximately $0.63^\circ$. We first focus on charge neutrality. Near $\deps=0$, electronic correlations open a small gap, which closes with a moderate applied displacement field\cite{han_correlated_2024}. For $|\deps|>$ \qty{0.3}{\volt\per\nano\meter}, a single-particle gap grows with increasing $\dfield$. Away from charge neutrality, vertical incompressible states correspond to integer fillings of the moir\'{e} superlattice. \figstext{} \ref{fig:fig1}c and \ref{fig:fig1}d show the width of the $\nu=1$ gap at \qty{0}{\tesla} for both signs of the displacement field as extracted from the compressibility data. The gap is significantly larger for positive $\dfield$, consistent with the stronger lateral localization of electrons at the moir\'{e} interface. Similarly, we observe almost no gap at $\nu=4$ for negative $\dfield$, indicating a much weaker moir\'{e} potential experienced by electrons than at positive $\dfield$ (\figtext{} \ref{fig:fig1}b). Beyond the gaps at integer filling, the qualitative differences in the map for either sign of the displacement field indicate that the moir\'{e} bandstructure differs significantly between the two cases. From here on, we focus on the moir\'{e}-proximal positive $\dfield$ limit.

\begin{figure*}
\includegraphics[width=\textwidth]{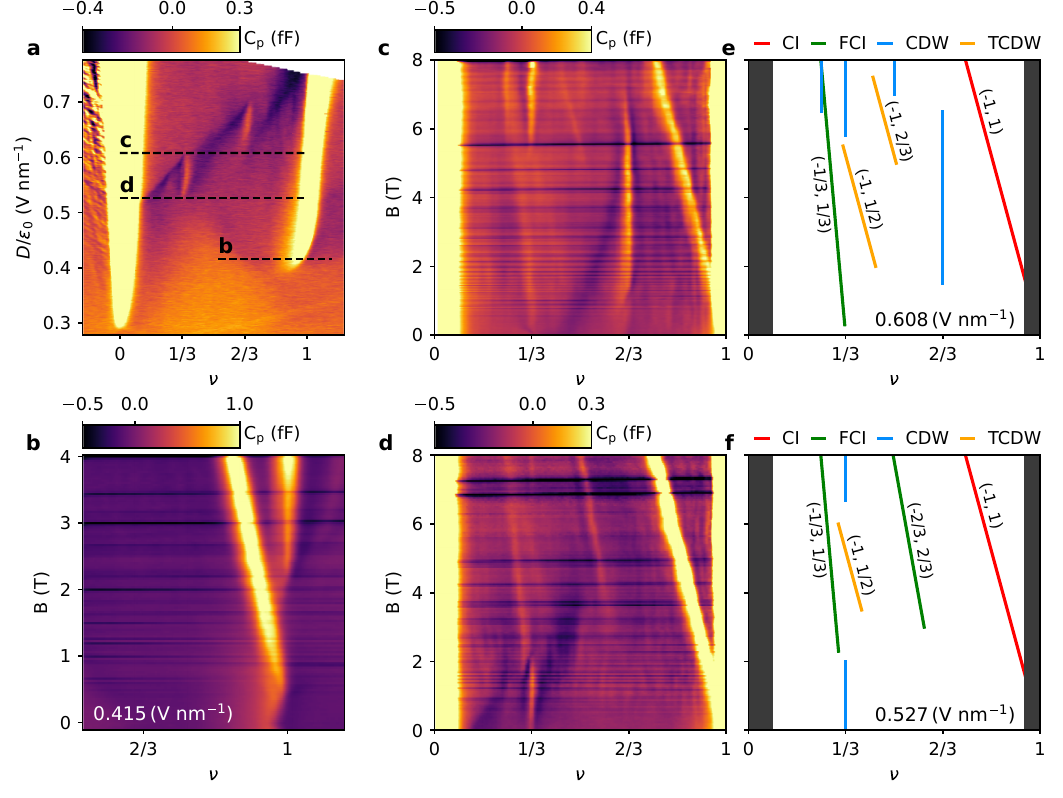}
\caption{\label{fig:fig2} \textbf{Magnetic field dependence of correlated topological states.} \textbf{a,} High-resolution map of the compressibility between $\nu=0$ and $\nu=1$ on the moir\'{e}-proximal side, corresponding to the boxed region in \figtext{} \ref{fig:fig1}b. Dashed lines correspond to the Landau fans in \textbf{b}-\textbf{d}. \textbf{b,} Magnetic field dependence of the $\nu=1$ state taken at $\deps=$ \qty{0.415}{\volt\per\nano\meter}. \textbf{c-f,} Compressibility Landau fans taken at $\deps=$ \qty{0.608}{\volt\per\nano\meter} (\textbf{c}) and $\deps=$ \qty{0.527}{\volt\per\nano\meter} (\textbf{d}), with their schematic representations shown in \textbf{e} and \textbf{f}, respectively. The horizontal bands in all Landau fans are due to cyclotron gaps in the graphite gates. Incompressible states are colored according to their topological classification: red for Chern insulators, green for fractional Chern insulators, yellow for topological charge density waves, and blue for trivial charge density waves.}
\end{figure*}

\section{Correlations and Topology}
In \figtext{} \ref{fig:fig2}a, we examine the electronic compressibility within the boxed region in \figtext{} \ref{fig:fig1}b. Beyond $\deps=$ \qty{0.4}{\volt\per\nano\meter}, the system develops a gap at filling factor $\nu=1$, and we measure enhanced compressibility between $\nu=0$ and $\nu=1$. Within this region, we observe a narrow stripe of negative compressibility\cite{bello_density_1981,tanatar_ground_1989,eisenstein_negative_1992} (where more charge enters the graphene than would be required to perfectly screen the applied ac excitation) moving diagonally from charge neutrality at $\deps=$ \qty{0.5}{\volt\per\nano\meter} to the $\nu=1$ gap at $\deps=$ \qty{0.75}{\volt\per\nano\meter}. This stripe contains two prominent charge density waves (CDW) at filling factors $\nu=1/3$ and $\nu=2/3$ along with weaker incompressible features developing around $\nu=1/4$ and $\nu=3/4$. Together, these phenomena indicate strong electronic correlations likely originating from a flat moir\'{e} band and a strong superlattice potential. In addition, as evidenced by the shift in density of the negative compressibility feature with $\dfield$, the displacement field enables fine control over the disperson of the moir\'{e} bands. To determine the topology of the $\nu=1$ state, we study the density dependence of the $\nu=1$ gap as a function of applied magnetic field (\figtext{} \ref{fig:fig2}b). We observe a $\mathcal{C}=-1$ state developing around \qty{0.5}{\tesla}, while a weaker trivial insulator only emerges above \qty{2}{\tesla}. Intriguingly, we measure a Chern number $\mathcal{C}=+1$ of opposite sign for the moir\'{e}-distant $\nu=1$ state\cite{tan_parent_2024} (Extended Data \figtext{} \ref{fig:topo_fan_both}), consistent with prior work in R5G-hBN\cite{lu_fractional_2024}. 

The strong correlations and intrinsic topology present in the moir\'{e}-proximal limit are two key ingredients for the observation of low-field fractional Chern insulators (FCI). A strong lattice potential can also stabilize topological charge density waves (TCDW), also referred to as ``symmetry-broken Chern insulators'', with coexisting topological order and broken translational symmetry\cite{wang_evidence_2015}. To search for these phenomena, we analyze the evolution of the compressibility within this band as a function of magnetic field. In the Hofstadter picture\cite{hofstadter_energy_1976,thouless_quantized_1982}, gapped states evolve with field according to $n_e=tn_\Phi+sn_0$, where $n_\Phi$ is the magnetic flux density per unit cell and $n_0$ is the unit cell density. The St\v{r}eda formula\cite{streda_theory_1982} then connects the slopes of these gaps to their associated Hall conductances through $\sigma_{xy}=te^2/h$. \figstext{} \ref{fig:fig2}c and \ref{fig:fig2}d are Landau fans taken along the two lines indicated in \figtext{} \ref{fig:fig2}a. These fans display a variety of topological states that we characterize according to their inverse slope $t$ and zero-field superlattice filling factor $s$.

\begin{figure*}
\includegraphics[width=\textwidth]{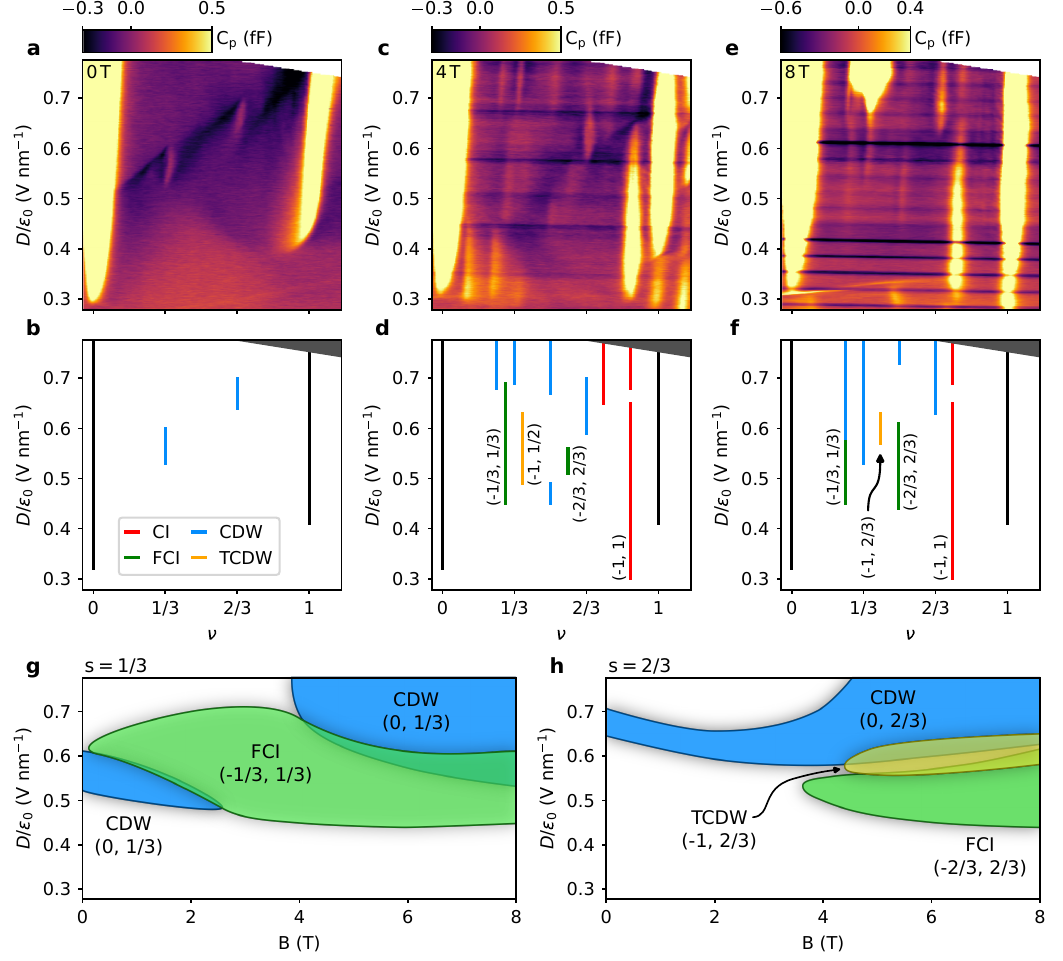}
\caption{\label{fig:fig3} \textbf{Displacement field-driven topological phase transitions.} \textbf{a-f,} Penetration capacitance maps between $\nu=0$ and $\nu=1$ as a function of density and displacement field at \qty{0}{\tesla} (\textbf{a}), \qty{4}{\tesla} (\textbf{c}), and \qty{8}{\tesla} (\textbf{e}), corresponding to the boxed region in \figtext{} \ref{fig:fig1}b. Their schematic representations are shown in $\textbf{b}$, $\textbf{d}$, and $\textbf{f}$, respectively. The horizontal bands are due to cyclotron gaps in the graphite gates. \textbf{d-f)} Schematic representation of the incompressible states observed in \textbf{a}, \textbf{b}, and \textbf{c}, respectively. States are colored according to their topological classification. \textbf{g,h,} Phase diagrams in magnetic and displacement field of correlated states emerging out of $\nu=1/3$ (\textbf{g}) and $\nu=2/3$ (\textbf{h}). The numbers in parentheses label $(t, s)$ for each state according to the Hofstadter model. Although the states overlap in displacement field, the individual phases occur at different densities for $B\neq0$.}
\end{figure*}

We begin by noting the presence of the integer CI $(t, s)=(-1,1)$ in both fans, although it is somewhat weaker at larger displacement field. FCI states occurring within this Chern band are expected to evolve with field as $(-\nu,\nu)$, where $0<\nu<1$. Indeed, we observe two such states, $(-1/3,1/3)$ and $(-2/3,2/3)$, in \figtext{} \ref{fig:fig2}d. In this map, we observe a transition between the 1/3 FCI and a trivial CDW state below \qty{2}{\tesla}. As we increase the displacement field, this transition is pushed to lower magnetic field (\figtext{} \ref{fig:fig2}c). The persistence of the FCI down to \qty{0.2}{\tesla} demonstrates that the moir\'{e} conduction bands have favorable quantum geometry even in the presence of a strong superlattice potential. The FCI to CDW phase transition can be induced either through tuning the magnetic field or the displacement field. Thus, the nature of the correlated ground state is quite sensitive to external control parameters, making this system a useful platform for studying topological phase transitions. Finally, we identify additional states with integer $t$ but fractional $s$; namely, $(-1,1/2)$ and $(-1,2/3)$, the aforementioned TCDWs. These states likely arise from commensurate charge density waves that distort the superlattice, doubling or tripling its unit cell. One of the filled subbands arising from this distortion then inherits the Chern number of the parent band\cite{xie_fractional_2021}.

The magnetic field dependence reveals a variety of fractional correlated phases. However, as is evident from the differences between \figstext{} \ref{fig:fig2}c and \ref{fig:fig2}d, these states are quite sensitive to displacement field. We generically expect that changing the displacement field will alter both the band dispersion and the Berry curvature distribution, leading to topological phase transitions. To that end, we now study the compressibility between $\nu=0$ and $\nu=1$ as a function of displacement field at different magnetic fields. We start with the zero-field case as a reference (\figstext{} \ref{fig:fig3}a and \ref{fig:fig3}b), again noting the two prominent CDW states. Next, we turn to the \qty{4}{\tesla} case (\figstext{} \ref{fig:fig3}c and \ref{fig:fig3}d). We observe the $\mathcal{C}=-1$ state with $(t, s)=(-1, 1)$ starting at $\deps=$ \qty{0.3}{\volt\per\nano\meter}. As we increase the displacement field beyond $\deps=$ \qty{0.65}{\volt\per\nano\meter}, the state terminates abruptly before reappearing very weakly above $\deps=$ \qty{0.7}{\volt\per\nano\meter}. Simultaneously, we find that the topological states also disappear above $\deps=$ \qty{0.65}{\volt\per\nano\meter}, with trivial CDWs taking their place. This pattern repeats at \qty{8}{\tesla} (\figstext{} \ref{fig:fig3}e and \ref{fig:fig3}f), with the FCI states vanishing along with the integer CI state at large $\dfield$ only to be replaced by CDWs. To summarize, we map out the displacement field-dependent ground state as a function of magnetic field for states with $s=1/3$ and $s=2/3$ (\figstext{} 3g and 3i, respectively). 

\begin{figure*}
\includegraphics[width=\textwidth]{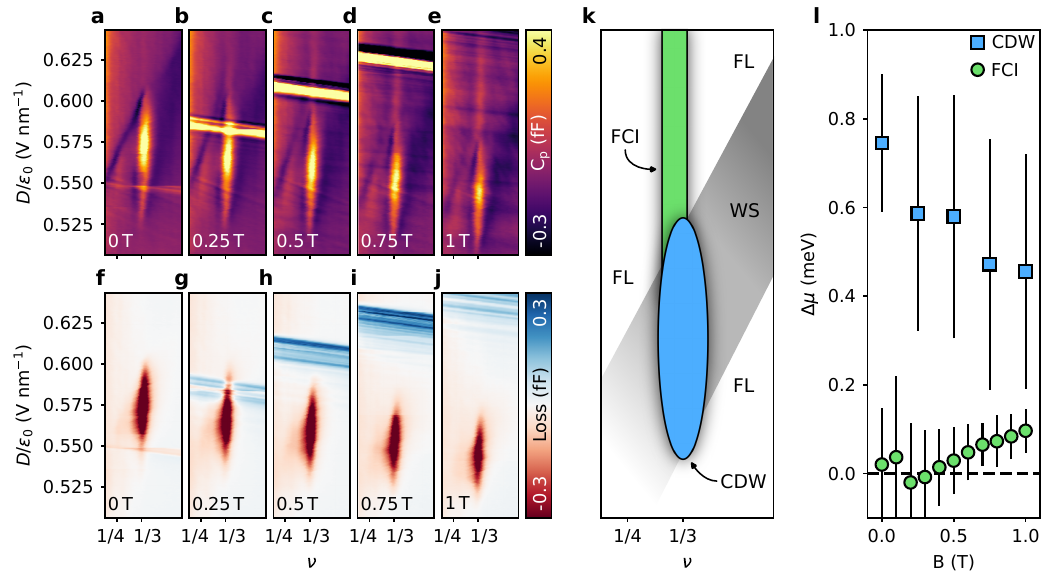}
\caption{\label{fig:fig4} \textbf{CDW-FCI transition.} \textbf{a-e,} Evolution of the $\nu=1/3$ CDW-FCI transition at low magnetic field, measured in a second device with the same twist angle. The bright horizontal feature is a measurement artifact which occurs at a fixed voltage on the back gate. Similar measurements on the primary device do not reproduce the very faint vertical bright line above the CDW at \qty{0}{\tesla} but do replicate the FCI at higher fields (Extended Data \figtext{} \ref{fig:extra_fci_d1}). \textbf{f-j,} Corresponding out-of-phase data for the plots in \textbf{a}-\textbf{e}. \textbf{k,} Schematic representation of the compressibility data, showing the locations of the Fermi liquid (FL), Wigner solid (WS), CDW, and FCI. \textbf{l,} Widths of the CDW and FCI gaps as extracted from the compressibility data. For the CDW, we show the maximum values of the gap widths in \textbf{a-e}. For the FCI, we show the mean width of the gap between 0.618 and \qty{0.640}{\volt\per\nano\meter} as determined from a different measurement on the same device (Extended Data \figtext{} \ref{fig:extra_fci_d2}).}
\end{figure*}

\section{CDW-FCI transition}
The competition between the CDW and FCI states at $\nu=1/3$ depends sensitively on the displacement field. To investigate this transition, we measure the compressibility at this filling factor over a narrow range of $\dfield$ (\figstext{} \ref{fig:fig4}a-e). We also extract the energy gaps of both correlated states as a function of magnetic field (\figtext{} \ref{fig:fig4}l). In addition, we display a schematic phase diagram of this region in \figtext{} \ref{fig:fig4}k. We first differentiate between two regimes away from $\nu=1/3$. In the upper-left and lower-right quadrants of the map, the system forms a highly-compressible Fermi liquid. Adjacent to the CDW, strong electronic correlations produce the stripe of negative compressibility discussed above. Corresponding transport measurements show an enhanced in-plane resistivity along this stripe (Extended Data \figtext{} \ref{fig:cdw_transport}). We thus attribute this state to an incommensurate Wigner solid or CDW pinned by defects. Since the density of the Wigner solid can change continuously, however, the phase still appears highly compressible. At $\nu=1/3$, on the other hand, the CDW is locked to the superlattice, generating a lattice distortion, and the system becomes highly incompressible. The loss component (\figstext{} 4f-j), which here increases with the graphene's in-plane resistance, shows a striking difference between the two correlated states. The CDW produces a sharp peak in the loss; in contrast, there is no corresponding feature for the FCI. This is consistent with the real-space picture of these states. The CDW is locked in place and cannot move or slide easily, making it difficult to charge the sample on each ac cycle. The more delicate FCI, while incompressible, has lower resistance since it is not pinned to a lattice distortion.

\section{Conclusions}
What accounts for the sensitivity of the correlated ground state to the displacement field? Rhombohedral multilayer graphene without a moir\'{e} superlattice already has a strongly displacement field-dependent bandstructure. At the $K$ points, the bottom of the conduction band flattens and then becomes dispersive again with increasing $\dfield$\cite{zhang_nearly_2019,dong_theory_2023}. With the addition of a superlattice potential, moir\'{e} minibands develop out of the flat region around the $K$ points. These minibands inherit the displacement field dependence of the original unfolded bands, thus allowing us to modulate their bandwidths by tuning $\dfield$. Electronic interactions may then amplify this effect by further restructuring the minibands. On the moir\'{e}-distant side, this picture is consistent with the FQAHE/FCIs occurring over a narrow range of displacement field\cite{lu_fractional_2024}. However, on the moir\'{e}-proximal side, the superlattice potential is strong enough to stabilize numerous CDWs in addition to FCI states.

Our observations stand in contrast to topological phase transitions seen in other graphene moir\'{e} systems\cite{spanton_observation_2018,xie_fractional_2021}. In this experiment, the displacement field plays an outsized role in modulating the moir\'{e} bandwidth and Berry curvature, providing an additional experimental control for tuning between different correlated ground states. Our results demonstrate that intrinsic band topology and strong electronic correlations in R5G-hBN persist in a strong superlattice potential. In particular, we establish R5G-hBN as a unique platform to study FCIs in both the moir\'{e}-proximal and moir\'{e}-distant limit, providing a bridge between the recently-reported FQAHE under a weak moir\'{e} potential and FCIs in other systems where the superlattice potential plays a dominant role. 

Further opportunities for bandstructure engineering in this system include a systematic study of the effect of twist angle on the intricate correlated phase diagram. Also, recent theoretical work suggests that the moir\'{e} band topology may survive even as the superlattice potential becomes vanishingly weak\cite{zhou_fractional_2023,dong_anomalous_2023}. This limit could be probed experimentally by imprinting a superlattce potential onto the pentalayer with a separate moir\'{e} heterostructure or patterned substrate\cite{zhou_fractional_2023,xu_creation_2021,ghorashi_topological_2023,zhang_engineering_2024,kim_electrostatic_2024,wang_band_2024}.  

\section*{Methods}
\textbf{Device fabrication.} The pentalayer graphene and hBN flakes were prepared by mechanical exfoliation onto SiO$_2$–Si substrates. The rhombohedral domains of pentalayer graphene were identified and confirmed using IR camera, near-field infrared microscopy, and Raman spectroscopy and isolated by cutting with a femtosecond laser. The van der Waals heterostructure was made following a dry transfer procedure. We picked up the top hBN, graphite, middle hBN and the pentalayer graphene using polypropylene carbonate film and landed it on a prepared bottom stack consisting of an hBN and graphite bottom gate. We aligned the long straight edge of graphene to that of hBN to nearly zero degrees to create a large moir\'{e} superlattice. The device was then etched into a multiterminal structure using standard e-beam lithography and reactive-ion etching. We deposited Cr–Au for electrical connections to the source, drain and gate electrodes.

\textbf{Compressibility meausrements.} All compressibility measurements in this work were performed in an Oxford Instruments Heliox $^{3}$He refrigerator with a base temperature of \qty{300}{\milli\kelvin}. The sample impedance is measured against a $\sim\,$\qty{25}{\femto\farad} reference in an impedance bridge circuit (Extended Data \figtext{} \ref{fig:full_circuit_diagram}). At the beginning of each measurement, independent ac excitaitons are applied to the sample and reference to null the signal on the balance point of the bridge. As the sample impedance changes, the new impedance can be computed from the resulting off-balance voltage on the balance point. In order to reliably measure the off-balance signal, a cryogenic amplifier\cite{steele_imaging_2006} is placed at the balance point of the bridge, which significantly reduces the output impedance of the circuit. Without this amplifier, the signal would be lost to the comparatively large parasitic capacitance of the coaxial cabling on the output line. At room temperature, the signal is recovered with a Stanford Research Systems SR865a lock-in amplifier. The ac excitations applied to the sample have an r.m.s.\@ amplitude of \qty{10}{\milli\volt} and a frequency of either \qty{10}{\kilo\hertz} (\figtext{} \ref{fig:fig1}, \figtext{} \ref{fig:fig2}a, and \figtext{} \ref{fig:fig3}) or \qty{150}{\kilo\hertz} (\figstext{} \ref{fig:fig2}b-d and \figtext{} 4).

\begin{acknowledgments}
We acknowledge helpful discussions with Liang Fu, Senthil Todadri, and Trithep Devakul. We thank Jackson Butler for a careful reading of the manuscript. Compressibility measurements were supported
by the STC Center for Integrated Quantum Materials, NSF grant no. DMR-1231319. L.J. acknowledges support from a Sloan Fellowship. S.A.\@ was supported by the NSF Graduate Research Fellowship under grant no.\@ 1122374.
\end{acknowledgments}

%\nocite{*}

\bibliography{main}% Produces the bibliography via BibTeX.

\clearpage % or \newpage?
\onecolumngrid
\appendix

%% Extended data -- start
\renewcommand\figurename{Extended Data Fig.}
\renewcommand\tablename{Extended Data Table}
\setcounter{figure}{0}
\setcounter{table}{0}

\begin{figure*}
\includegraphics[width=\textwidth]{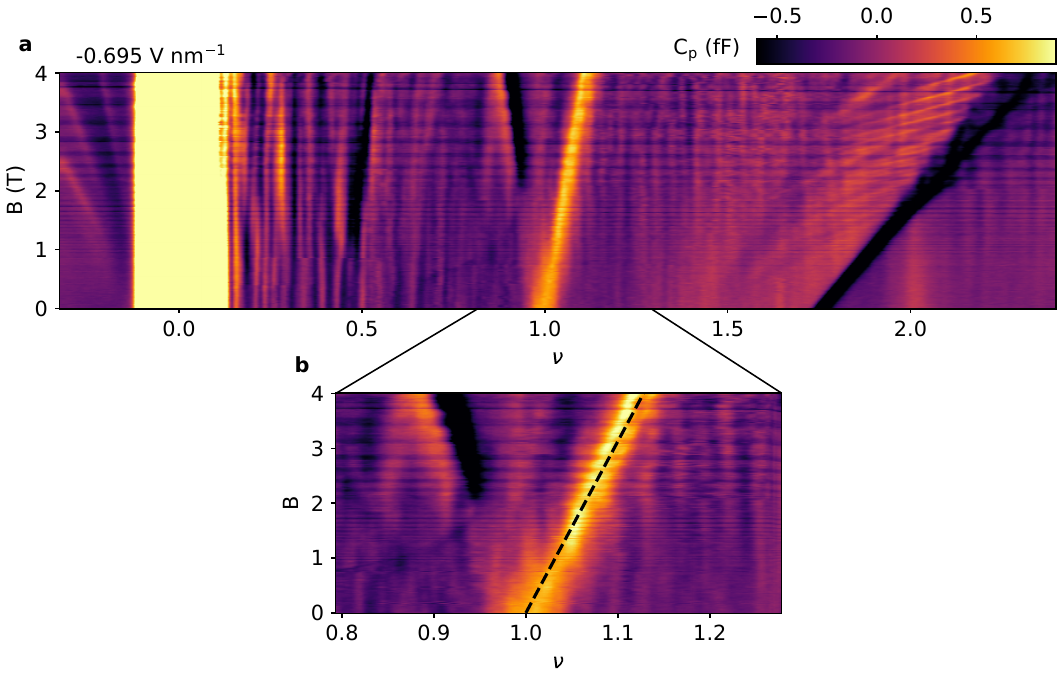}
\caption{\label{fig:topo_fan_both} \textbf{Topology on the moir\'{e}-distant side.} \textbf{a,} Magnetic field dependence of the compressibility measured at $\deps=$ \qty{-0.695}{\volt\per\nano\meter} and \textbf{b,} a zoomed-in view on $\nu=1$. The dashed line in \textbf{b} traces out the the expected evolution of a gapped state with $\mathcal{C}=+1$.}
\end{figure*}

\begin{figure*}
\includegraphics[width=0.667\textwidth]{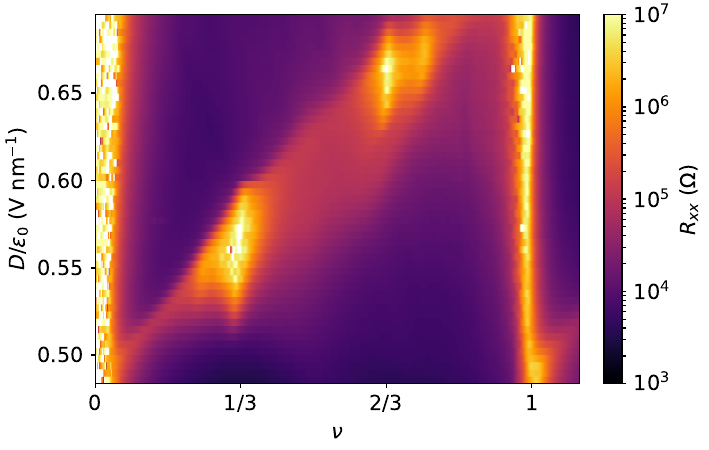}
\caption{\label{fig:cdw_transport} \textbf{Tranpsort measurement between $\nu=0$ and $\nu=1$ on the moir\'{e}-proximal side.} The resistive diagonal stripe through the center of the map corresponds to the stripe of negative compressibility discussed in the main text. Prominent CDW states appear at $\nu=1/3$ and $\nu=2/3$, with weaker states forming at $\nu=1/4$ and $\nu=3/4$.}
\end{figure*}

\begin{figure*}
\includegraphics[width=\textwidth]{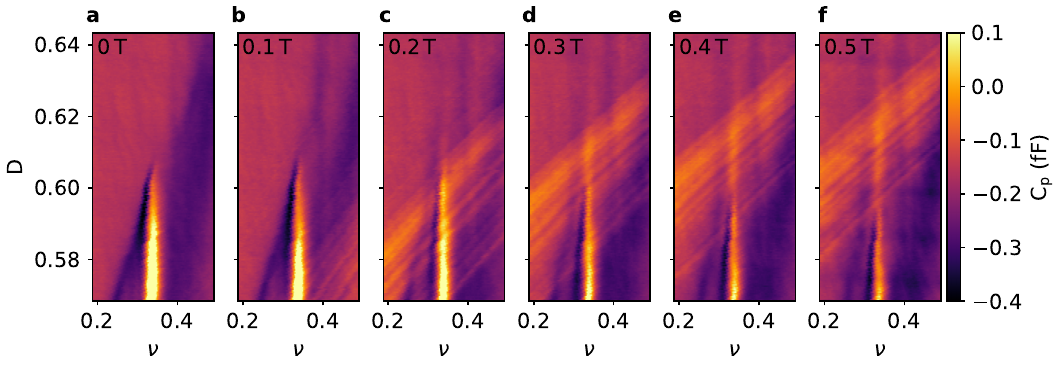}
\caption{\label{fig:extra_fci_d1} \textbf{FCI-CDW transition in Device DI.} Evolution of the $\nu=1/3$ CDW-FCI transition at low magnetic field, measured in device DI. The FCI is barely visible above the CDW starting at \qty{0.2}{\tesla} (\textbf{c}). In contrast to device DII, we observe no bright line above the CDW at \qty{0}{\tesla}.}
\end{figure*}

\begin{figure*}
\includegraphics[width=\textwidth]{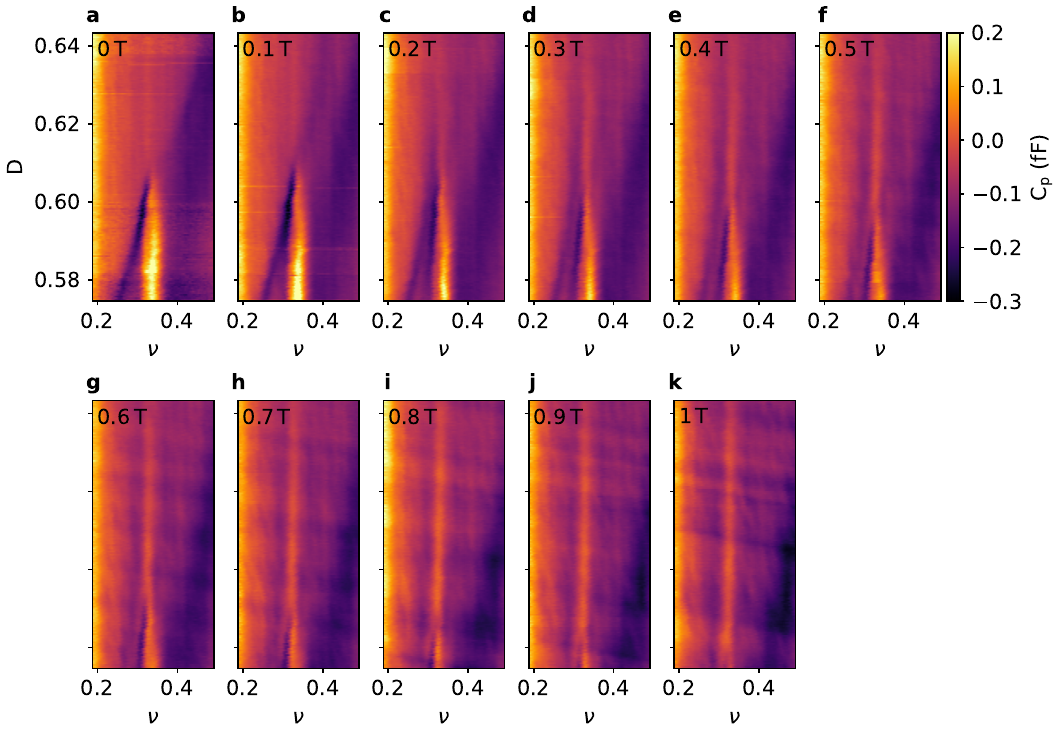}
\caption{\label{fig:extra_fci_d2} \textbf{Additional measurements of the FCI-CDW transition in Device DII.} A second set of measurements corresponding to those presented in \ref{fig:fig4}\textbf{a}-\textbf{e}, now taken over a narrower range in displacement field and more finely spaced in magnetic field. The FCI emerges out of the tip of the CDW starting at \qty{0.2}{\tesla}. This dataset was used to compute the FCI gaps in \figtext{} \ref{fig:fig4}l.}
\end{figure*}

\begin{figure*}
\includegraphics[width=\textwidth]{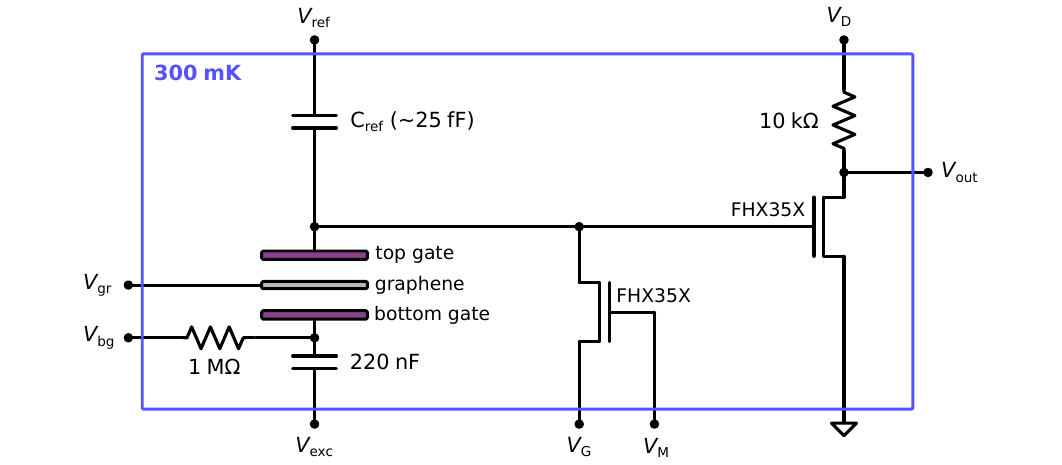}
\caption{\label{fig:full_circuit_diagram} \textbf{Impedance bridge and cryogenic amplifier circuit for compressibility measurements}}
\end{figure*}

\begin{figure*}
\includegraphics[width=\textwidth]{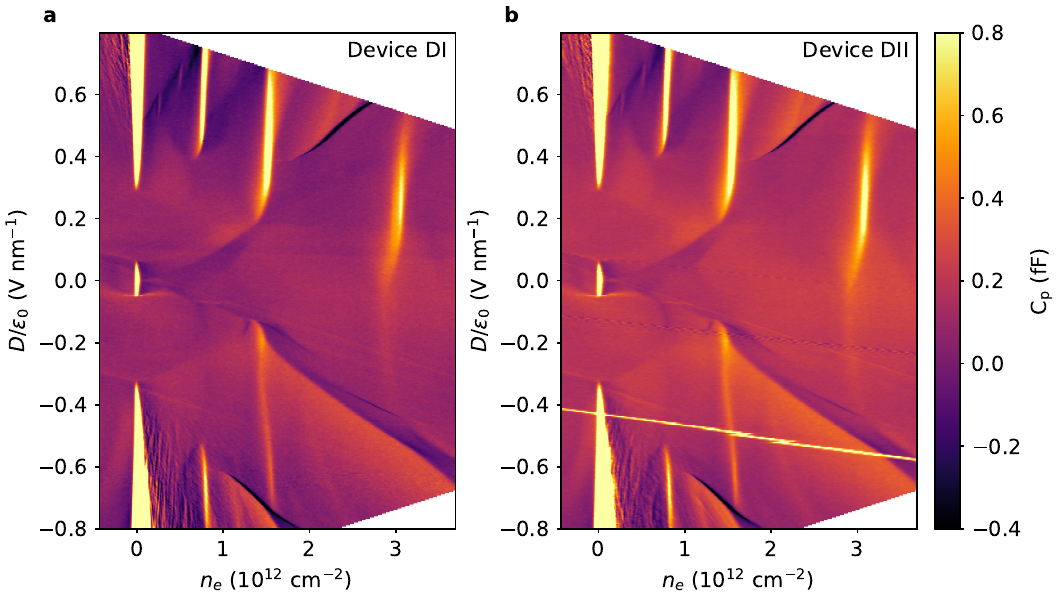}
\caption{\label{fig:large_map_comparison} \textbf{Comparison of compressibility phase diagrams for DI and DII.} Device DI (\textbf{a}) was used for the data shown in \figstext{} \ref{fig:fig1}, \ref{fig:fig2}, and \ref{fig:fig3}. Device DII (\textbf{b}) was used for the data shown in \figtext{} \ref{fig:fig4}, including the FCI and CDW gaps. }
\end{figure*}

\clearpage
\begin{figure*}
\includegraphics[width=\textwidth]{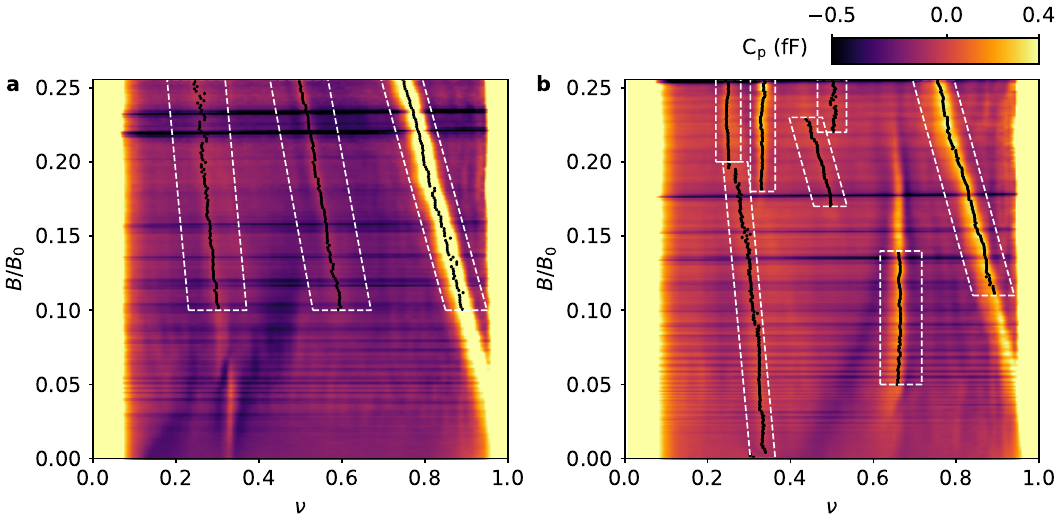}
\caption{\label{fig:fan_fits} \textbf{Linear fits to incompressible states in Landau fans.} The same two Landau fans displayed in \figtext{} \ref{fig:fig2}, taken at $\deps=$ \qty{0.527}{\volt\per\nano\meter} (\textbf{a}) and $\deps=$ \qty{0.608}{\volt\per\nano\meter} (\textbf{b}). Black dots mark the least compressible point at each magnetic field within the corresponding white dashed box. The vertical axis corresponds to magnetic flux quanta per superlattice unit cell. Fits to the incompressible states are presented in Extended Data Table \ref{tab:slope_fits}.}
\end{figure*}

\begin{table}
    \centering
    \begin{tabular}{c|c|cc|cc}
        Fan & State & $t$ (calculated) & $s$ (calculated) & $t$ (assigned) & $s$ (assigned)\\
        \hline
         1 & FCI & -0.31(2) & 0.33(1) & -1/3 & 1/3 \\
         1 & FCI & -0.63(1) & 0.66(1) & -2/3 & 2/3 \\
         1 & CI  & -0.93(2) & 0.99(1) & -1   & 1   \\
         % not showing this fan
         %\hline
         %2 & FCI & -0.35(3) & 0.34(1) & -1/3 & 1/3 \\
         %2 & CI  & -0.94(3) & 1.01(1) & -1   & 1   \\
         %2 & TCDW& -0.98(6) & 0.67(1) & -1   & 2/3 \\
         \hline
         2 & FCI & -0.35(2) & 0.34(1) & -1/3 & 1/3 \\
         2 & CI  & -0.98(1) & 1.00(1) & -1   & 1   \\
         2 & TCDW& -1.03(4) & 0.68(1) & -1   & 2/3 \\
         2 & CDW & -0.02(4) & 0.25(1) & 0    & 1/4 \\
         2 & CDW & 0.07(3)  & 0.32(1) & 0    & 1/3 \\
         2 & CDW & -0.02(13)& 0.51(3) & 0    & 1/2 \\
         2 & CDW & 0.03(2)  & 0.66(1) & 0    & 2/3 \\
         \hline
    \end{tabular}
    \caption{Computed and assigned $t$ and $s$ values for incompressible states in the Landau fans shown in Extended Data \figtext{} \ref{fig:fan_fits}. The values correspond to linear fits through each set of points in those fans. Errors correspond to 95\% confidence intervals for the calculated parameters.}
    \label{tab:slope_fits}
\end{table}

%% Extended data -- end

\clearpage

%% supplement -- start

\renewcommand\figurename{Fig.}
\renewcommand\tablename{Table}
\renewcommand{\thefigure}{S\arabic{figure}}
\renewcommand{\thetable}{S\arabic{table}}
\setcounter{figure}{0}
\setcounter{table}{0}

\section*{Supplementary Information}
\setcounter{equation}{0}
\subsection*{Determining gap sizes from the quantum capacitance}
Thermodynamic gaps can be obtained by integrating the quantum capacitance $c_q=e^2\partial n/\partial\mu$ over the carrier density as
\begin{equation}
    \Delta\mu=e^2\int dn\,c_q^{-1}
\end{equation}
In practice, we control the density indirectly through the top and bottom gate voltages $V_t$ and $V_b$. The assumption that $\delta n$ is proportional to $\delta V$ breaks down when the electronic compressibility is sufficiently low and the quantum capacitance contributes significantly to the total capacitance. Thus, we make the substitution
\begin{equation}
    e\delta n\ \ \rightarrow\ \ \left(\frac{1}{c_t}+\frac{1}{c_q}\right)^{-1}\delta V_t\,\,+\,\,\left(\frac{1}{c_b}+\frac{1}{c_q}\right)^{-1}\delta V_b
\end{equation}
in the integral for $\Delta\mu$, which yields
\begin{equation}
    \Delta\mu=e\int dV_t\,\left(1+c_q/c_t\right)^{-1}\,+\,e\int dV_b\,\left(1+c_q/c_b\right)^{-1}
\end{equation}
\subsection*{Determining the quantum capacitance from the measured impedance ratio}
The impedance bridge does not measure $c_q$ directly. Instead, it determines the ratio of the impedance of the reference $Z_r$ to that of the sample $Z_s$, which is a function of the quantum capacitance as well as the sheet resistance $R_s$ of the graphene (in practice, the reference is purely capacitive, so $Z_r=1/i\omega C_r$). Using a distributed circuit model for the device\cite{dultz_thermodynamic_2000} derived below, we can approximate this ratio as 
\begin{equation}
    \frac{Z_r}{Z_s(c_q,R_s)}=\frac{A}{C_r}\left(\frac{c_tc_b}{c_t+c_b}\right)\left[1-\frac{c_q}{c_q+c_t+c_b}\frac{\tanh(\alpha/2)}{\alpha/2}\right]+\frac{C_\text{back}}{C_r}
\end{equation}
where
\begin{equation}
    \alpha=\sqrt{i\omega AR_s\,\frac{c_q(c_t+c_b)}{c_q+c_t+c_b}}
\end{equation}
We recast this expression in terms of experimentally measurable quantities. We define $\chi_\text{band}$ as the value of $Z_r/Z_s$ when the sample is highly compressible: $c_q\gg c_t, c_b$, $R_s\ll 1/i\omega A(c_t+c_b)$, and $\alpha\rightarrow 0$. Likewise, we define $\chi_\text{gap}$ as the value of $Z_r/Z_s$ when $R_s\gg 1/i\omega A(c_t+c_b)$ and $\alpha\rightarrow\infty$. We can then write
\begin{equation}
    \frac{Z_r}{Z_s(c_q,R_s)}=\left(\chi_\text{gap}-\chi_\text{band}\right)\left[1-\frac{c_q}{c_q+c_t+c_b}\frac{\tanh(\alpha/2)}{\alpha/2}\right]+\chi_\text{band}
\end{equation}
We now have two coupled implicit equations for $c_q$ and $R_s$ in terms of the real and imaginary components of $Z_r/Z_s$ (measured on the lock-in amplifier) and the geometric capacitances $c_t$ and $c_b$ (determined from the Landau fans). Finally, we numerically invert the equations for each data point to obtain the quantum capacitance. The device area $A$ only appears as a product with $R_s$ and is therefore irrelevant in computing $c_q$.

\subsection*{Determining $\chi_\text{gap}$}
\begin{figure*}
\includegraphics[width=\textwidth]{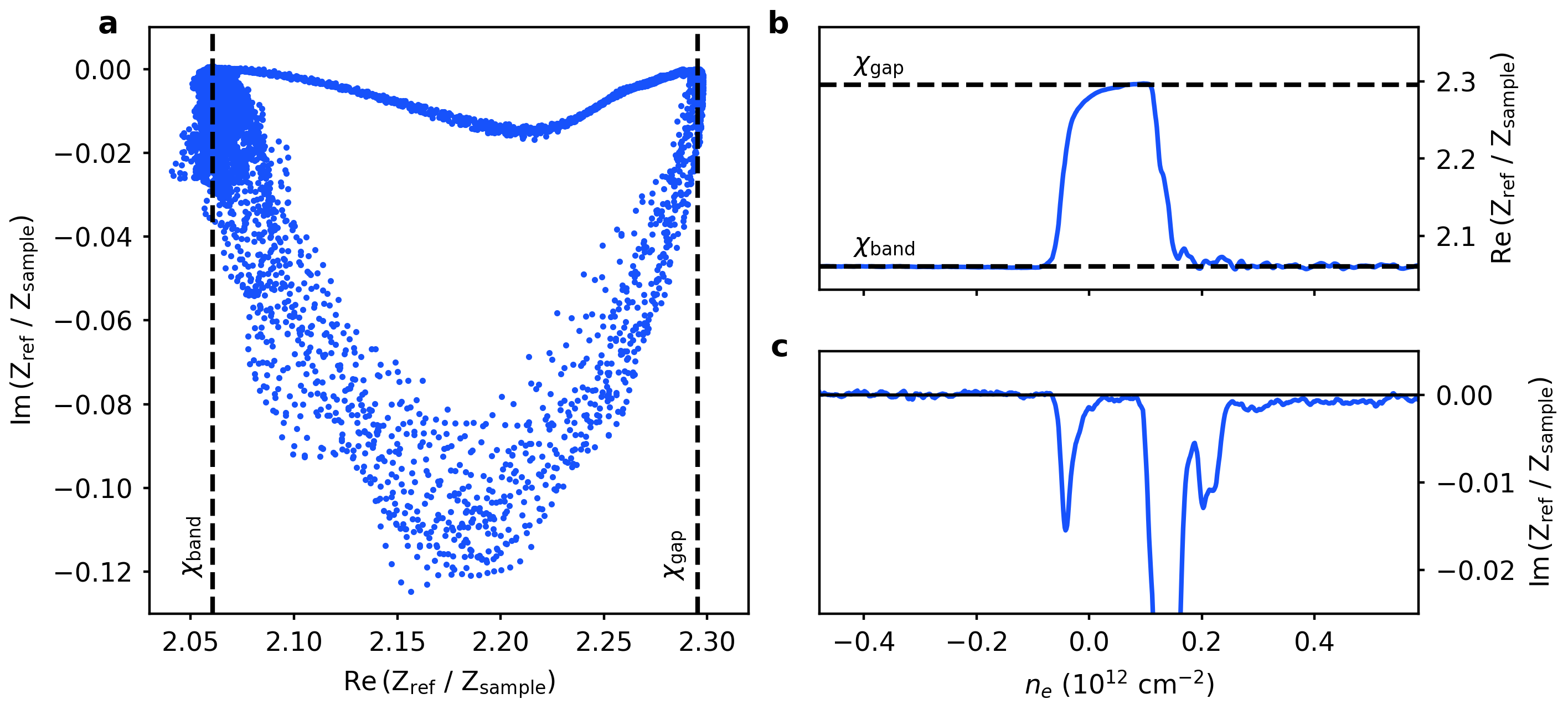}% Here is how to import EPS art
\caption{\label{fig:phasemap} a) Scatter plot of the real and imaginary components of the complex impedance ratio $Z_\text{r}/Z_\text{s}$ measured at large displacement field. b) $\text{Re}(Z_r/Z_s)$ and c) $\text{Im}(Z_r/Z_s)$ measured as a function of electron density, showing the single-particle band gap at charge neutrality.} 
\end{figure*}
To determine an accurate value for $\chi_g$, we need to measure the capacitance in the incompressible limit. \figtext{} \ref{fig:phasemap} shows a linecut of the real (capacitive) and imaginary (resistive) components of the impedance ratio as the density is swept through charge neutrality at large displacement field. Deep in the band, the sample fully screens the applied excitation. The signal is small and purely capacitive as we measure only the shunt capacitance. As we approach the band edge, the loss grows substantially with the in-plane resistance of the sample. Once we are in the gap, however, the sample cannot charge at all, and the signal again becomes purely capacitve. In this limit, we measure the sum of the shunt capacitance and the geometric capacitance of the device. 

\figtext{} \ref{fig:phasemap} shows the corresponding scatter plot of the real and imaginary components of the signal. Points cluster along the real axis at $\chi_b$ and $\chi_g$. To determine $\chi_g$, we bin the data along this axis and find the peak of the distribution. Then, we obtain the uncertainty $\delta\chi_g$ as the difference between this peak value and the maximum value of $\text{Re}\left(Z_r/Z_s\right)$ occurring in the dataset.

\subsection*{Determining $\chi_\text{band}$}
\begin{figure*}
\includegraphics[width=\textwidth]{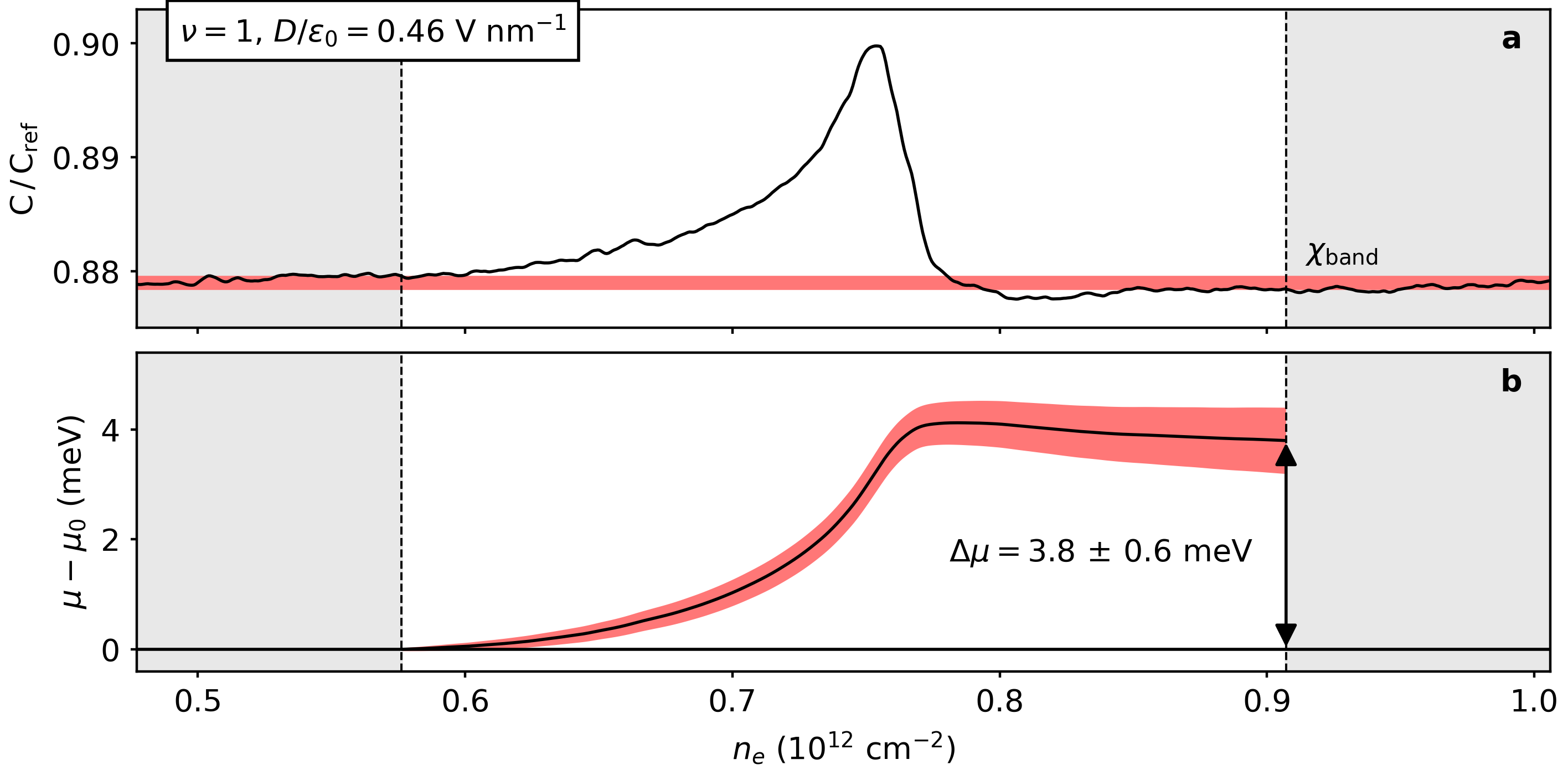}% Here is how to import EPS art
\caption{\label{fig:gap_methodology} a) Measured capacitance ratio $C_s/C_r$ as a function of electron density across filling factor $\nu=1$. The red band indicates the standard deviation of $C_s/C_r$ in the grey shaded regions on either side of the plot. b) The inverse compressibility $(\partial n/\partial\mu)^{-1}$ integrated over the gap. Error accumulates in the integral due to the uncertainty in $\chi_\text{band}$.}
\end{figure*}
Determining an appropriate value for $\chi_b$ is a more difficult task, for two reasons. First, the gaps we wish to integrate over exist on top of a background of finite and/or non-constant compressibility. Second, noise levels in this background can be significant when compared to the heights of the incompressible peaks. Although we could use the dataset in \figtext{} \ref{fig:phasemap} to make a global estimate for $\chi_b$, we choose to use a local background for each integral. \figtext{} \ref{fig:gap_methodology} is a schematic representation of this procedure for the moir\'{e}-proximal $\nu=1$ gap. We split the dataset into two regions: one containing the gap itself, and the other containing two narrow highly-compressible regions on either side of the gap. At each displacement field, we obtain reasonable values for $\chi_b$ and $\delta\chi_b$ by taking the mean and standard deviation of the measured capacitance ratio $\chi$ in the compressible regions. We then use these values in the integrals for the gap width and its error. \figtext{} \ref{fig:gap_methodology}a shows the result of this procedure. The grey regions on either side are used for determining $\chi_b$, while the white region in the center is used in the integral. The noise and drift in the background is captured within the red band, representing the uncertainty in $\chi_b$. This uncertainty leads to error accumulating within the subsequent integral for $\Delta\mu$ (\figtext{} \ref{fig:gap_methodology}b). 

\subsection*{Low-frequency approximation}
In the low-frequency limit, $\alpha\rightarrow0$, and the impedance ratio simplifies to
\begin{equation}
    \frac{C_s}{C_r}=\frac{A}{C_r}\left(\frac{1}{c_t}+\frac{1}{c_b}+\frac{c_q}{c_tc_b}\right)^{-1}+\frac{C_\text{back}}{C_r}
\end{equation}
where the expression in parentheses is just the inverse penetration capacitance per unit area. From here on, we use $\chi$ to refer to the measured impedance ratio $C_s/C_r$. We can rewrite this equation in terms of experimentally determined quantities by defining $\chi_\text{band}$ as the value of $\chi$ when $c_q\rightarrow\infty$ and the system is highly compressible. Likewise, $\chi_\text{gap}$ denotes the value of $\chi$ when $c_q=0$ and the system is fully gapped. After some rearranging, we obtain
\begin{equation}
    \frac{1}{c_q}=\frac{1}{e^2}\frac{\partial\mu}{\partial n}=\left(\frac{\chi-\chi_\text{band}}{\chi_\text{gap}-\chi}\right)\frac{1}{c_t+c_b}
\end{equation}
\subsection*{Determining the error in the integrand}
Now that we have an analytic low-frequency approximation for $c_q$, we can propagate the errors in $\chi_\text{band}$ and $\chi_\text{gap}$ through the integrands. We assume that the error in the geometric capacitance is significantly smaller and neglect it here. We consider the integrand for the top gate voltage,
\begin{equation}
m_t=\left[1+\frac{c_t+c_b}{c_t}\left(\frac{\chi_\text{gap}-\chi}{\chi-\chi_\text{band}}\right)\right]^{-1}
\end{equation}
The error is then given by
\begin{equation}
    \delta m_t=\sqrt{\left(\frac{\partial m_t}{\partial \chi_\text{band}}\right)^2\delta\chi_\text{band}^2+\left(\frac{\partial m_t}{\partial \chi_\text{gap}}\right)^2\delta\chi_\text{gap}^2}
\end{equation}
with partial derivatives
\begin{align}
    \frac{\partial m_t}{\partial \chi_\text{band}}&=\frac{m_t(m_t-1)}{\chi-\chi_\text{band}}\\[1.5ex]
    \frac{\partial m_t}{\partial \chi_\text{gap}}&=\frac{m_t(m_t-1)}{\chi_\text{gap}-\chi}
\end{align}
We then obtain the error in $\Delta\mu$ as 
\begin{equation}
    \delta(\Delta\mu)=e\int dV_t\,(\delta m_t)\,+\,e\int dV_b\,(\delta m_b)
\end{equation}

\subsection*{Derivation of the impedance ratio}
Consider a simple parallel-plate capacitor. Between the two plates is a third layer, say, a graphene flake, with sheet resistance $R_s$ and quantum capacitance per unit area $c_q=e^2\partial n/\partial\mu$ which is grounded through an ohmic contact. A voltage excitation $\varepsilon(t)=\varepsilon_0 e^{i\omega t}$ is applied to the bottom plate. What is the current $I(t)$ that flows onto the top plate? If the graphene were absent, the current in the circuit would simply be given by
\begin{equation}
    I(t)=i\omega c_g\varepsilon(t)\iint d\mathbf{x}
\end{equation}
where $c_g=c_tc_b/(c_t+c_b)$ is the geometric capacitance per unit area between the top and bottom plates and the integral is over the area of the system. However, the electrostatic potential $\phi(\mathbf{x}, t)=\phi(\mathbf{x}) e^{i\omega t}$ that develops on the graphene partially screens the field from the bottom plate, so we instead have 
\begin{equation}
    I(t)=i\omega c_g\iint d\mathbf{x}\left[\varepsilon(t)-\phi(\mathbf{x},t)\right]
\end{equation}
or, after dividing by $\varepsilon(t)$ to obtain the admittance,
\begin{equation}
    \frac{1}{Z}=i\omega c_g\iint d\mathbf{x}\left[1-\frac{1}{\varepsilon_0}\phi(\mathbf{x})\right]
\end{equation}
We begin by recasting the integral in terms of the total potential $V(\mathbf{x}, t)$ and current density $\mathbf{K}(\mathbf{x}, t)$ on the graphene. The total potential is a sum of electrostatic and chemical contributions:
\begin{equation}
    V(\mathbf{x}, t)=\phi(\mathbf{x}, t)+\frac{1}{e}\mu(\mathbf{x}, t)
\end{equation}
After differentiating both sides with respect to time, we obtain
\begin{equation}
    i\omega V=i\omega\phi+\left(\frac{1}{e^2}\frac{\partial\mu}{\partial n}\right)\frac{\partial\sigma}{\partial t}
\end{equation}
where $\sigma$ is the sheet charge density on the graphene and the quantity in parentheses is just the inverse quantum capacitance per unit area $c_q$. Here we assume that the applied voltage excitation is small enough that $c_q$ is nearly constant across the graphene. Invoking charge continuity $\partial_t\sigma+\nabla\cdot\mathbf{K}=0$ and rearranging the expression, we have
\begin{equation}
    \phi(\mathbf{x}, t)=V(\mathbf{x}, t)+\frac{1}{i\omega c_q}\nabla\cdot\mathbf{K}(\mathbf{x}, t)
\end{equation}
and the integral for the admittance becomes
\begin{equation}
    \frac{1}{Z}=i\omega c_g\iint d\mathbf{x}\left[1-\frac{1}{\varepsilon_0}V(\mathbf{x})-\frac{1}{i\omega c_q\varepsilon_0}\nabla\cdot\mathbf{K}(\mathbf{x})\right]
\end{equation}
To complete the problem, we must determine $V(\mathbf{x})$ and $\mathbf{K}(\mathbf{x})$. We again differentiate the total potential with respect to time, now writing the result in a different form: 
\begin{equation}
    \frac{\partial V}{\partial t}=\left(\frac{\partial\phi}{\partial\sigma}+\frac{1}{e^2}\frac{\partial\mu}{\partial n}\right)\frac{\partial\sigma}{\partial t}
\end{equation}
The two terms in parentheses are the inverse geometric and quantum capacitances per unit area, respectively. Again invoking charge continuity, we can rewrite this as an equation relating $V$ and $\mathbf{K}$:
\begin{equation}
    \nabla\cdot\mathbf{K}(\mathbf{x}) = -i\omega\left(\frac{c_q(c_t+c_b)}{c_q+c_t+c_b}\right)V(\mathbf{x})
\end{equation}
Ohm's law gives us a second equation relating $V$ and $\mathbf{K}$ through the sheet resistance of the graphene:
\begin{equation}
    \nabla V(\mathbf{x})=-R_s\mathbf{K}(\mathbf{x})
\end{equation}
Rearranging, we obtain separate equations for $V$ and $\mathbf{K}$:
\begin{equation}
    \nabla^2V=\gamma^2V,\ \ \nabla\left(\nabla\cdot\mathbf{K}\right)=\gamma^2\mathbf{K},\ \ \gamma^2=i\omega R_s\left(\frac{c_q(c_t+c_b)}{c_q+c_t+c_b}\right)
\end{equation}
Once we specify the sample geometry (namely, the shape of the graphene flake and the locations of the contacts), we can solve these two equations and feed the solutions back into the admittance integral. 

The sample studied in this work is a transport device with multiple contacts along its edges, which we shorted to a common ground during capacitance measurements. Thus, we model the device as a one-dimensional transmission line of length $L$ with grounded ohmic contacts on both ends. Instead of sheet resistance and capacitance per unit area, both $R$ and $c$ are now per unit length. The equations for voltage and current on the graphene become
\begin{equation}
    \frac{\partial^2V}{\partial x^2}=\gamma^2V,\ \ \frac{\partial^2I}{\partial x^2}=\gamma^2I
\end{equation}
The voltage along the line drops according to
\begin{equation}
    \frac{\partial V}{\partial x}=RI(x)
\end{equation}
Solving these three equations, we obtain the general solutions
\begin{align*}
    I(x)&=a_1e^{\gamma x}+a_2e^{-\gamma x}\\
    V(x)&=Z_c\left(-a_1e^{\gamma x}+a_2e^{-\gamma x}\right)
\end{align*}
where $Z_c$ is the characteristic impedance of the line:
\begin{equation}
    Z_c=\sqrt{\frac{R}{i\omega}\left(\frac{c_q+c_t+c_b}{c_q(c_t+c_b)}\right)}
\end{equation}
To determine $a_1$ and $a_2$, we require that $V(x)=\varepsilon_0$ at both ends of the line ($x=0$ and $x=L$). We can then plug $V(x)$ and $I(x)$ back into the admittance integral to obtain
\begin{equation}
    \frac{1}{Z}=i\omega c_gL\left[1-\frac{c_q}{c_q+c_t+c_b}\frac{\tanh(\gamma L/2)}{\gamma L/2}\right]
\end{equation}
Generalizing to a real device, we can replace the length with an area $A$ and add a background capacitance to arrive at the expression for the impedance ratio used above.
\end{document}